\documentclass[conference]{IEEEtran}
\IEEEoverridecommandlockouts

\usepackage{amsmath,amssymb,amsfonts}
\usepackage{algorithmic}
\usepackage{graphicx}
\usepackage{textcomp}
\usepackage{xcolor}
\usepackage{comment}

\usepackage{gensymb}
\usepackage{color}
\usepackage{graphicx}  
\usepackage{soul}
\usepackage{physics}

\usepackage[ruled,vlined,linesnumbered]{algorithm2e}
\usepackage[numbers,sort&compress,square]{natbib}
\usepackage{enumerate}
\usepackage[normalem]{ulem}
 
\usepackage{pdfpages}

\usepackage[margin=.72in]{geometry} 

\usepackage{multirow}
\usepackage[normalem]{ulem}
\usepackage{float}
\usepackage{scalerel}
\usepackage{tikz}
\usetikzlibrary{svg.path}

\usepackage{algorithmic}

\usepackage{subcaption}

\usepackage{url}

\def\BibTeX{{\rm B\kern-.05em{\sc i\kern-.025em b}\kern-.08em
    T\kern-.1667em\lower.7ex\hbox{E}\kern-.125emX}}
 
\usepackage[font=small,skip=0pt,labelsep=period,justification=justified]{caption}
    
\usepackage[normalem]{ulem}

\begin{document}

\title{An ADMM-based MIQP platform for the EV aggregation management}
\author{Shahab Afshar, Shailesh Wasti and Vahid Disfani}

\author{\IEEEauthorblockN{ Shahab Afshar}
\IEEEauthorblockA{\textit{ConnectSmart Research Laboratory} \\
\textit{University of Tennessee at Chattanooga}\\
Chattanooga, TN, USA \\
shahab-afshar@mocs.utc.edu}
\and
\IEEEauthorblockN{ Shailesh Wasti}
\IEEEauthorblockA{\textit{ConnectSmart Research Laboratory} \\
\textit{University of Tennessee at Chattanooga}\\
Chattanooga, TN, USA \\
shailesh-wasti@mocs.utc.edu }
\and
\IEEEauthorblockN{ Vahid Disfani}
\IEEEauthorblockA{\textit{ConnectSmart Research Laboratory} \\
\textit{University of Tennessee at Chattanooga}\\
Chattanooga, TN, USA \\
vahid-disfani@mocs.utc.edu }
}
\maketitle
\begin{abstract}
Electric vehicle (EV) aggregation can significantly influence the EVs charging/discharging behavior. In this paper, we use a distributed algorithm based on the alternating direction method of multipliers (ADMM) to coordinate EV charging and discharging procedures for EVs with vehicle-to-grid (V2G) capabilities. 
The optimization model is formulated as a mixed-integer quadratic programming (MIQP) problem to consider the efficiency of EV batteries and different energy prices in both charging and discharging processes. 
Numerical tests using real-world data confirms that the implemented method allows obtaining both the electric vehicle aggregator (EVA) and individual EV goals while considering the power grid and each EV constraints. Moreover, we show the significant impact of our model on the final demand profile and computation time.

\end{abstract}

\IEEEpeerreviewmaketitle

\let\thefootnote\relax\footnotetext{© 2020 IEEE. Personal use of this material is permitted. Permission from IEEE must be obtained for all other uses, in any current or future media, including reprinting/republishing this material for advertising or promotional purposes, creating new collective works, for resale or redistribution to servers or lists, or reuse of any copyrighted component of this work in other works.}

\let\thefootnote\relax\footnotetext{}
\let\thefootnote\relax\footnotetext{This paper has been accepted for presentation at the International Conference on Smart Grids and Energy Systems (SGES 2020), to be held in Perth, Australia from 23-26 November 2020.} 

\section{Introduction}

\IEEEPARstart{T}{he} expansion of the electric vehicle fleet in the last decade has been remarkable. It is expected that the global energy demand for EV will increase from 20 billion kWh in 2020 to about 280 billion kWh in 2030 \cite{finance2019electric}. This unprecedented growth of EVs has the potential to avert the catastrophic impacts of climate change. But the challenges of the integration of EVs at the grid-edge can't be overlooked \cite{afshar2020literature}. Uncontrolled penetration of EVs in the distribution grid might have adverse impacts on the power grid \cite{mohsenzadeh2018optimal}. Against this backdrop, designing a control infrastructure for a large electric vehicle fleet is one of the main challenges of the EV aggregator. EVA, usually an existing entity or a profit-aimed utility, bridges EVs and distribution system operators for the energy management and provides mutual advantages for both \cite{wei2015charging}. 

A centralized architecture for energy management with various functionalities has been extensively reported in the literature. \cite{clement2009impact} implements a centralized EV fleet control model that reduces voltage deviations and lower power losses by shaving peak load. To address the uncertainty of driving pattern, \cite{zheng2018novel} realized a centralized and stochastic coordination method to flatten the elevated peak loads arising from EV charging. However, with the significant growth of EVs and the sheer volume of data associated with it, a traditional centralized framework is challenging in both computation and communication front. Distributed paradigms, on the other hand, have sought a lot of attention recently because of their robust, scalable, and privacy-preserving characteristics \cite{kia2019tutorial}.

Distributed algorithms can be classified based on consensus theory and (sub)gradient method. Gradient-based algorithms are well established distributed algorithms \cite{yang2019survey, disfani2015optimization}. A gradient descent algorithm for constrained optimization referred to as "Consensus and Innovation" is modeled in \cite{mohammadi2016fully} as a cooperative charging problem for plug-in EVs. Authors in \cite{gan2013optimal} implemented a projected gradient descent (PGD), and in \cite{liu2019decentralized} a shrunken primal-dual subgradient (SPDS) algorithm for controlling EVs to solve the load variance minimization problem. 

Among the distributed algorithms, an alternating method of multiplier (ADMM) algorithm based on dual decomposition is extensively implemented for different applications because of its capabilities to handle large-scale problems and to support not strictly convex function. \cite{boyd2011distributed,mota2011proof,wasti2020distributed}. Authors in \cite{vaya2014decentralized} flattened the load curve, considering the uncertainty in driving patterns. The power supply cost is minimized in \cite{zhang2017scalable} while respecting substation capacity limitations and voltage regulation. Since the replacement battery cost is one of the major costs of an EV, the reduction in the life span of the battery due to continuous charging and discharging process is considered in \cite{carli2017decentralized,rivera2017distributed,mohiti2019decentralized,khaki2019hierarchical}.

Battery replacement cost is one of the major costs in an EV, which naturally inspires to investigate the optimal operation of batteries.  
Previously, we developed an upper-level state of charge (SOC) based management system in \cite{miao2013soc}. In \cite{carli2017decentralized}, battery degradation and congestion in the power grid are considered while investigating the optimal scheduling of the EV fleets. 
To the best of the authors’ knowledge, previous research works do not formulate a distributed EV charging management problem with individual charging and discharging constraints, and positive and negative power exchange energy tariffs.

In this paper, we formulate an ADMM \textit{exchange problem} between individual EVs and EVA, where each agent (any EV or EVA) has its own objective function and constraint sets. As the optimization problem is strongly coupled with a Lagrange variable, the feasible set of any agent influences the set of other agents. As a case study, we set the objective of load variance minimization (LVM) and charging cost minimization (CCM) for EVA, and battery degradation function for EVs. As that the optimization problem between EVs is fully decoupled, Evs can have their own battery degradation function. The contribution of this paper is summarized in three points as follows:

 \begin{itemize}
 
\item  Reformulating the non-linear EV control framework as an  ADMM-based MIQP problem to guarantee the global optimality of the solution. 
 
\item Creating a comprehensive model for battery operation in the optimization problem including energy constraints and charging/discharging efficiencies.

\item Considering separate tariffs for positive and negative power exchange with the grid, and different charging and discharging power constraints for the EVA.

\end{itemize}

The rest of the paper is structured as follows. Section II provides the mathematical formulation of the general objective function and  presents the EVA and EVs objective functions and constraints. Section III reformulates the optimization problem based on the ADMM method. Simulations and result analyses are provided in Section IV. Section V concludes the paper and envisions future work.
\section{Optimization Problem}

This paper formulates a coordinated EV charging control infrastructure for a finite time horizon  $t~\in \mathcal{T}= ~\{1,\ldots ,T \}$.
In this platform, aggregators and EVs seek different objectives in their optimization problem. 
Assume that one EVA defines the optimal scheduling of $N$ EVs, where EVA and EVs may have different objective functions. Therefore, the scheduling optimization can be represented as follows \cite{rivera2017distributed}: 
\begin{subequations}
\begin{align} 
{\text {minimize}}&~F_{a}(x_{a}) + \gamma \sum \nolimits _{i=1}^{N} F_{i}(x_{i}) \\
 \mbox {subject to}~ &x_{a} = \sum \nolimits _{i=1}^{N} x_{i}\label{power_balance_constrant} \\ 
& x_{a} \in \mathbb {X}_{a} \\
& x_{i} \in \mathbb {X}_{i}; \quad i=1,\ldots ,N 
\end{align}
\label{EVA_original}
\end{subequations}
where $x_i$ and $F_{i}(x_{i})$ denote the power demand and the objective function of EV $i$,  $x_a$ and $F_{a}(x_{a})$ denotes those of EVA, and the weighting factor $\gamma>0$ makes a trade-off between these objectives \cite{rivera2017distributed}. In this paper, we also study the impacts of different values of $\gamma$ on the optimization objectives. The constraint \eqref{power_balance_constrant} also ensures the power balance equation between EVA and EVs, and $\mathbb {X}_{i}$ and $\mathbb {X}_{a}$ denote the feasible sets of EV $i$ and aggregator's decision variables. The comprehensive optimization models for both entities as described below.
\subsection{EV Optimization Problem}
The degradation of the lifespan of the battery is modeled as a quadratic polynomial function, which corresponds to minimizing the (dis)charging cycle of the battery. The objective function and constraints of each EV, thus, can be mathematically written as follows:

\begin{subequations}
\setlength{\belowdisplayskip}{1.5pt} \setlength{\belowdisplayshortskip}{0pt}
\setlength{\abovedisplayskip}{1.5pt} \setlength{\abovedisplayshortskip}{0pt}
\begin{align}
& \text {minimize} \; \alpha_{i}\left\|x_{i}\right\|_{2}^{2} \label{eq: EV obj}
\\
& E_{i}^{0}+ \sum_{t=1}^{T}(p^{ch}_{i,t}\eta_{i}^{ch}-p^{{dis}}_{i,t}/\eta_{i}^{dis})/ m=R_{i} \label{F:ev_C1} \\
& x_{i,t}=p^{ch}_{i,t}-p^{dis}_{i,t} \label{F:ev_C2} \\
& u^{ch}_{i,t}+u^{dis}_{i,t} \leq A_{i,t} \label{F:ev_C3} \\
& 0 \leq p^{ch}_{i,t} \leq \overline p^{ch}_{i,t}u^{ch}_{i} \label{F:ev_C4} \\
& 0 \leq p^{dis}_{i,t} \leq \overline p^{dis}_{i,t}u^{dis}_{i} \label{F:ev_C5}  \\
& E_{i,t+1}=E^{0}_{i}+\frac{1}{m}\sum^{T}_{t=1}(\eta_{i}^{ch}p^{ch}_{i,t}-p^{dis}_{i,t}/\eta_{i}^{dis}) \label{F:energy_cons1} \\
& \underline E_{i,t} \leq E_{i,t} \leq \overline E_{i,t} \label{F:energy_cons2} \end{align}
\label{F:EVobj} \end{subequations}
where $\alpha_i$ is the battery degradation coefficient and $x_i$ is the charging profile of $\text{EV}_{i}$.
In the set of constraints of each EV, $m$ is the number of time steps in each hour; $R_{i}$ is $\text{EV}_{i}$ charging requirements; charging and discharging variables are denoted by $p^{ch}_{i,t}$ and $p^{dis}_{i,t}$, and $\eta_{i}^{ch}$ and $\eta_{i}^{dis}$ stand for EVs' charging and discharging efficiency, respectively. Energy balance equation is presented in (\ref{F:ev_C1}), and (\ref{F:ev_C2}) decomposes the $x_{i}$ to charging and discharging variables. To do the decomposition, we used charging and discharging binary variables in (\ref{F:ev_C3}), which are denoted by $u^{ch}_{i,t}$ and $u^{dis}_{i,t}$, respectively. However, these new variables convert the QP to an MIQP optimization problem \cite{feizollahi2015large}. (\ref{F:ev_C3}) shows that EVs charging and discharging cannot be done simultaneously where $A_{i,t}$ stands for $\text{EV}_{i}$ connection status. We assumed that each EV knows its required energy and its connection status, which is a binary parameter. (\ref{F:ev_C4}) and (\ref{F:ev_C5}) are present the maximal charging and discharging constraints. Due to the technical limit of EVs batteries and its importance on the aggregated EVs load profile, these constraints are formulated in (\ref{F:energy_cons1}) and (\ref{F:energy_cons2}). The equality constraint expresses the EV battery energy balance at each time step and the inequality constraint presents the energy limits equation. 

ADMM-based platform solves the MIQP problem by capturing the discrete nature of the newly-added binary decision variables \cite{feizollahi2015large,takapoui2017alternating}. Moreover, the ADMM-based method is proven to ensure the convergence in a finite number of iterations \cite{feizollahi2015large}. Even with a linear objective function for EV and EVA optimization problems, we still face a quadratic optimization problem due to the quadratic term of the augmented Lagrangian functions.
\subsection{Aggregator Optimization Problem}

Two typical EVA objectives, load variance minimization and charging cost minimization, are specified in this section.

\subsubsection{Load variance minimization}

In order to help the power grid to control the demand profile, EVA uses (\ref{F:Load_var}) to do the peak shaving in peak hours and valley filling in off-peak hours. 
The load variance minimization problem can be formulated as follows \cite{rivera2017distributed}:
\begin{equation} \text {minimize} ~ \delta \left |{\left |{D+x_{a} }\right |}\right |_{2}^{2}
\label{F:Load_var}
\end{equation}
where $\delta$ is a scaling parameter used as an empirical value that changes the unit of the EVA objective function and the non-EV demand profile $D \in \mathbb {R}^{T}$, which is assumed to be known. 

\subsubsection{Charging cost minimization}
In order to modify customers' charging behavior and smooth the load profile, DN considers different electricity prices for peak and off-peak hours. The EVA objective here is minimizing the charging cost of each EV, according to this price. EVA charging cost minimization and it's constrains are formulated as follows:
\begin{subequations}
\setlength{\belowdisplayskip}{1.5pt} \setlength{\belowdisplayshortskip}{0pt}
\setlength{\abovedisplayskip}{1.5pt} \setlength{\abovedisplayshortskip}{0pt}
\begin{align}  & \text {minimize}~ (\sum^{T}_{t=1}(\pi^{ch}_{a,t}p^{ch}_{a,t}-\pi^{dis}_{a,t} p^{dis}_{a,t}) / m) \label{F:Charging_cost_minimization} \\ &
u^{ch}_{a,t}+u^{dis}_{a,t} <= 1 \label{F:Charging_cost_minimization_cons2}\\ &   
x_{a,t}=p^{ch}_{a,t}-p^{dis}_{a,t} \label{F:Charging_cost_minimization_cons1} \\ & 0 \leq p^{ch}_{a,t} \leq \overline p^{ch}_{a,t}u^{ch}_{a,t}\label{F:Charging_cost_minimization_cons3} \\ & 0 \leq p^{dis}_{a,t} \leq \overline p^{dis}_{a,t}u^{dis}_{a,t}\label{F:Charging_cost_minimization_cons4}\end{align}
\label{F:Charging_cost_minimization_cons}
\end{subequations} 

where $\pi^{t}_{ch}$ and $\pi^{t}_{dis}$ denote charging and discharging prices respectively. The aggregated charging and discharging variables are also denoted by $p^{ch}_{a,t}$ and $p^{dis}_{a,t}$.  It can be seen from (\ref{F:Charging_cost_minimization}) that we consider different tariffs for positive and negative power exchange with the power grid, which is necessary when EVs have vehicle to grid (V2G) discharging capability.

In (\ref{F:Charging_cost_minimization_cons2}), $u^{ch}_{a,t}$ and $u^{dis}_{a,t}$ are charging and discharging binary variables at any time $t\in\mathcal{T}$, used to decompose $x_{a}$ to charging and discharging variables in (\ref{F:Charging_cost_minimization_cons1}). (\ref{F:Charging_cost_minimization_cons3}) and (\ref{F:Charging_cost_minimization_cons4}) determine the grid exchange power limits for charging and discharging processes accordingly. In other words, feeder capacity constraints for (dis)charging are considered in these two equations, separately.

\section{ADMM-Based EV Aggregation Models}
Solving the optimization problem \eqref{EVA_original} can become unmanageable when several EV $i\in \{1, \ldots , N\}$ with individual charging profile $x_{i}=[ x_{i}(1),\ldots ,x_{i}(T)]^{\top}$ need to be considered. To resolve this concern, we develop an ADMM based method to decompose the optimization problem into smaller sub-problems between aggregators and EVs. 

\subsection{Problem Decomposition}
In the optimization problem \eqref{EVA_original}, the objective function and all the constraints except (\ref{power_balance_constrant}) can be directly decomposed between EVs and aggregator. 
If the constraint (\ref{power_balance_constrant}) is ignored, each EV can define its optimal charging profile locally $F_{i}(x_{i})$ from its own feasible set $\mathbb {X}_{i}$. The EVA also calculates the optimal charging and discharging power profiles $x_{a}=[ x_{a}(1),\ldots ,x_{a}(T)]^{\top}$ within its feasible set $\mathbb {X}_{a}$ to minimize $F_{a}(x_{a})$. 

To formulate the ADMM-based optimization algorithm, first,
the problem should be reformulated as an exchange optimization problem. Since EVA and EVs have separate sub-problems,
to simplify the notation, it is assumed that  $x_{0}=-x_a$, which redefines the aggregator as sub-problem 0. On the other hand, the EVs are represented by sub-problems $i=1,\ldots ,N$. The cost function of EVA and EV sub-problems are shown in (\ref{F:sub_eva}) and (\ref{F:sub_ev}), respectively.
\begin{equation} f_{0}(x_{0})=\begin{cases} F_{a}(-x_{0}) & \text {if} -x_{0} \in \mathbb {X}_{a}\\ \infty & \text {otherwise} \end{cases} \label{F:sub_eva}\end{equation}
\begin{equation} f_{i}(x_{i})=\begin{cases} \gamma F_{i}(x_{i}) & \text {if}~x_{i} \in \mathbb {X}_{i}\\ \infty & \text {otherwise}\\ \end{cases} \label{F:sub_ev} \end{equation}
Considering (\ref{F:sub_eva}) and (\ref{F:sub_ev}), the exchange problem can be defined as follows \cite{boyd2011distributed}:
\begin{align} {\text {minimize}}&\sum \nolimits _{i=0}^{N} \,f_{i}(x_{i}) \notag \\[3pt] \mbox {subject to}&\sum \nolimits _{i=0}^{N}x_{i} = 0. \label{f:exchangeproblem} \end{align}
To simplify notation, the optimization
variables of all sub-problems in (\ref{f:exchangeproblem}) are defined in $x = [x_{0}, x_{1}, \dots , x_{N}]^{\top}$.
\subsection{ADMM-Based Algorithm}
To define EVs' optimization functions, we borrow the exchange problem from the model proposed in  \cite{boyd2011distributed}. The exchange problem can be solved using ADMM.
\begin{align}
&x_{i}^{k+1}:=\underset{x_{i}}{\operatorname{argmin}}\left(f_{i}\left(x_{i}\right)+(\rho / 2)\left\|x_{i}-x_{i}^{k}+\bar{x}^{k}+\frac{\lambda^{k}}{\rho}\right\|_{2}^{2}\right) \notag \\
&\lambda^{k+1}:=\lambda^{k}+\rho\bar{x}^{k+1} \label{f:exchange}
\end{align}
where $\lambda^k$ denotes the Lagrange multiplier associated with the power balance constraint \eqref{power_balance_constrant} at iteration $k$, $\rho >1$ denotes the penalty parameter of the augmented term, and $\overline {x}^{k}$ denotes the average power mismatch of the entire platform at iteration $k$ calculated as
\begin{align} \overline {x}^{k+1} = \frac {1}{N+1} \big(x_{0}+\sum _{i=1}^{N} x_{i}^{k+1}\big) \label{F:General_UPDATE_ADMM}
\end{align} 

\begin{figure}
    \centering
    \includegraphics[width=1\columnwidth]{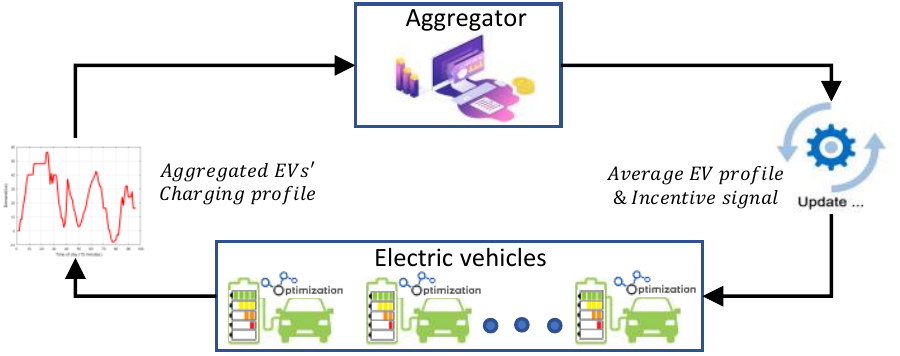}
    \caption{Schematic of the system}
    \label{fig:my_label1}
\end{figure}

 ADMM-based EV algorithm has three main steps in which EV and EVA optimize their objective functions in the exchange problem platform.  Fig.~\ref{fig:my_label1} shows the overall framework of EV ADMM problem. 
The ADMM platform leads to the optimal solution of the original problem through the following update processes at each iteration $k$:
\begin{align}
&x_{i}^{k+1} = \left\{ \begin{matrix} \text {argmin}& \gamma F_{i}(x_{i}) + \frac {\rho }{2} \norm{{x_{i}- x_{i}^{k} + \overline {x}^{k} + \frac{\lambda^{k}}{\rho}}}_{2}^{2} \\
\text{subject to}&x_{i} \in \mathbb {X}_{i} \end{matrix} \right\}
\label{F:General_EV_ADMM}\\
&x_{0}^{k+1} =\left\{ \begin{matrix}  \text {argmin}& F_{a}(-x_{0}) + \frac {\rho }{2} \norm{x_{0}- x_{0}^{k} + \overline {x}^{k+1} + \frac{\lambda^{k}}{\rho}}_{2}^{2} \\ \text{subject to}&-x_{0} \in \mathbb {X}_{\text{a}} \end{matrix} \right\}
\label{F:General_EVA_ADMM}
\end{align}
\begin{align}
&\lambda^{k+1} = \lambda^{k} + \rho\overline {x}^{k+1} \label{F:up_beta}
\end{align}

More specifically, each EV $i$ solves 
\begin{align} 
&x_{i}^{k+1} = \left\{ \begin{matrix} \text {argmin}& \gamma \alpha _{i} \norm{{x_{i} }}_{2}^{2} + \frac {\rho }{2} \norm{{x_{i}- x_{i}^{k} + \overline {x}^{k} + \frac{\lambda^{k}}{\rho}}}_{2}^{2} \\
\text{subject to}&x_{i} \in \mathbb {X}_{i} \end{matrix}\right\}
\end{align}
at each iteration $k$ to minimize the degradation impacts.  Similarly, the EVA optimization problem in (\ref{F:Load_var}) is reformulated according to (\ref{F:General_EV_ADMM}) as 
\begin{align} 
x_{0}^{k+1} = \frac {\rho }{\rho +2\delta } \left ({ x_{0}^{k}- \overline {x}^{k} - \frac{\lambda^{k}}{\rho} }\right ) + \frac {2\delta }{\rho +2\delta } D
\end{align}
and the EVA optimization problem in (\ref{F:Charging_cost_minimization}) is
reformulated based on (\ref{F:General_EVA_ADMM}) as 
\begin{align} 
&x_{0}^{k+1} =\left\{ \begin{matrix}  \text {argmin}& (\sum^{T}_{t=1}(\pi^{dis}_{a,t}p^{dis}_{a,t}-\pi^{cha}_{a,t} p^{cha}_{a,t}) / m) +\\
&\frac {\rho }{2} \norm{x_{0}- x_{0}^{k} + \overline {x}^{k+1} + \frac{\lambda^{k}}{\rho}}_{2}^{2} \\ \text{subject to}&-x_{0} \in \mathbb {X}_{\text{a}} \end{matrix} \right\}  
\end{align}

\subsection{Convergence Criteria}
To stop the back-and-force procedure between the EVA and EVs, we need appropriate convergence criteria. In our ADMM-based model, the primal feasibility $r^{k} \in \mathbb {R}^{T}$ and dual feasibility $s^{k}_{i} \in \mathbb {R}^{T}$ are specified the convergence criteria \cite{boyd2011distributed}.
\begin{align}\centering &r^{k}=\overline {x}^{k} \label{F:primal} \\ 
s^{k}_{i}=-\rho (N+1)& \left ({x^{k}_{i} - x^{k-1}_{i} + \left ({\overline {x}^{k-1} - \overline {x}^{k} }\right )}\right ) \label{F:dual} \\
 &||r^{k}||_{2}\leq\epsilon ^{p} \notag \\ &||s^{k}||_{2}\leq\epsilon ^{d}  
\label{F:epsilon}
\end{align}

Equation (\ref{F:primal}) and (\ref{F:dual}) show the primal and dual convergence criteria, respectively. While $s^{k}=[s^{k}_{1},\ldots ,s^{k}_{N}]^{\top}$, stopping criteria can be defined, considering $\epsilon ^{p}$ and $\epsilon ^{d}$ in (\ref{F:epsilon}).
\subsection{ADMM based Operation Framework}
The following ADMM-based operation framework displays a back-and-force procedure between the EVA and EVs using ADMM. According to its local constraints $\mathbb {X}_{i}$, each EV obtains its charging profile locally $F_{i}(x_{i})$ and sends this information to the EVA through a messaging system. 
EVA then calculates the summation of charging $x_{a}=[ x_{a}(1),\ldots ,x_{a}(T)]^{\top}$, for the global objective $F_{a}(x_{a})$ which is subjected to global constraints set $\mathbb {X}_{a}$. It is obvious that $x_{a}= \sum _{i=1}^{N} x_{i}$, through which the problems are coupled. Note that $\rho$ determines the speed of convergence and the stability status of the ADMM algorithm. Here, we select a value of $\rho$ within the stable region to guarantee convergence. See \cite{boyd2011distributed} for more details on selection of $\rho$.

\begin{algorithm}
    \begin{algorithmic}
    \footnotesize
   \STATE \textbf{Initialization:  $\lambda^{k}  , \: \overline  {x}^{k}$ and $\rho$}
   \STATE $k=0$
   
    \WHILE{(\ref{F:epsilon}) is not true}

      \FOR{$i=1:N$}
       \STATE Solve the EVs optimization problems using (\ref{F:General_EV_ADMM})

            \ENDFOR 

       \STATE Solve the EVA optimization problem using (\ref{F:General_EVA_ADMM})

        \STATE Update  $\overline {x}^{k+1}$ by (\ref{F:General_UPDATE_ADMM}) and $\lambda^{k+1}$ by (\ref{F:up_beta})

        \STATE Calculate $r^{k}$ and $s^{k}$ by (\ref{F:primal}) and (\ref{F:dual}) and check (\ref{F:epsilon})

      \STATE Send $\overline {x}^{k+1}$ and $\lambda^{k+1}$ to EVs
      \STATE   $k=k+1$
      \ENDWHILE
    \end{algorithmic}
\caption{ADMM-based Algorithm}
\vspace{5pt}
\label{Alg:ADMM}
\end{algorithm}
\section{Simulation results}

\subsection{Case study}
\begin{figure}[hpbt!]
    \centering
    \includegraphics[width=1\linewidth]{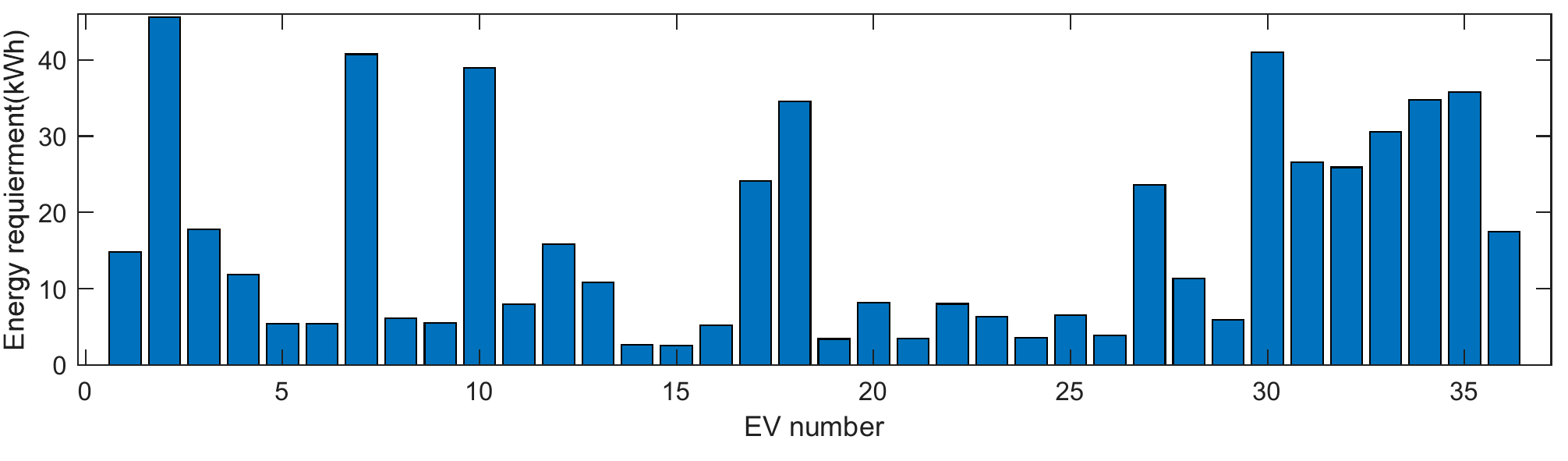}
    \caption{Charging requirement (kWh) }
    \label{fig:A}
\end{figure}

\begin{figure}[hpbt!]
    \centering
    \includegraphics[width=1\columnwidth]{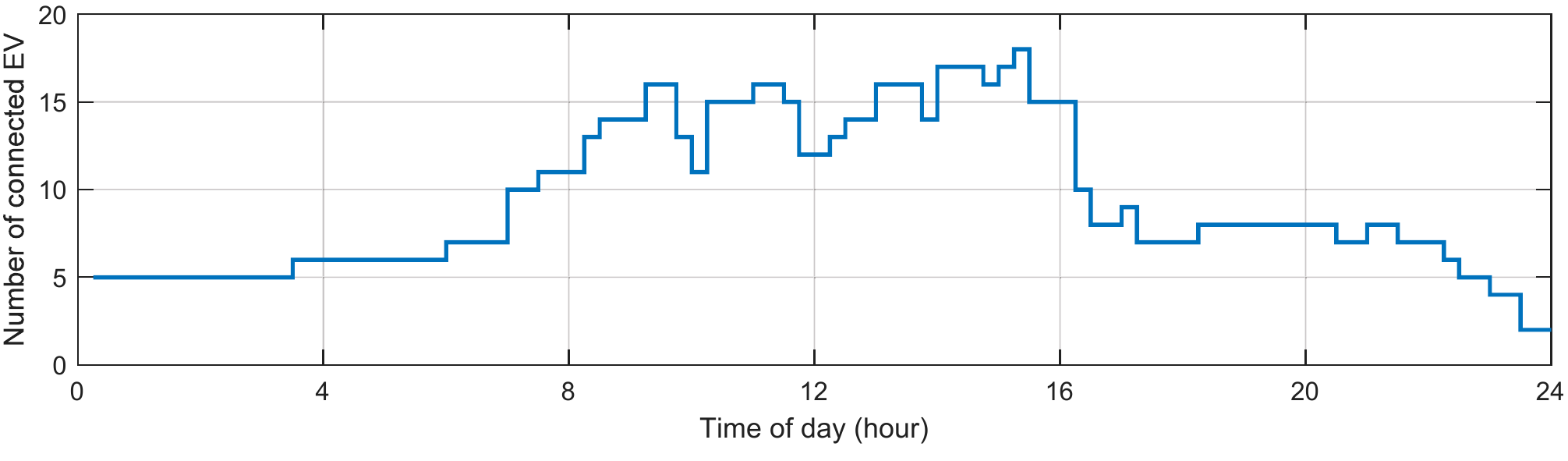}
    \caption {Connection status}
    \label{fig:R}
\end{figure}

To simulate this paper, we assume that the aggregator has access to the non-EV demand profile and the electricity price. The demand profile is assumed to be known and its data is extracted from the Munich distribution system operator website for every 15 minutes in a sample day \cite{munich}. For simulations, we employ CARTA's \cite{carta} data-set of real EV charging events in the city of Chattanooga on a sample day of March 2, 2020. This information are including the connection time of each EV, energy requirement, and the location of the charging stations for 36 charging events. We assumed that each EV can only have one charging event per day. It is worth pointing out that to the best of our knowledge, most of the previous documents considered full-time connectivity for all EVs, which is not a realistic assumption. Time of use (TOU) electricity price is also obtained from the southern California time of use rate plan \cite{tou}, where the electricity price is 14 cent/kWh for off-peak and 38 cent/kWh for peak hours. We assumed that the $\pi^{t}_{dis}$ is 40\% of $\pi^{t}_{ch}$. 
CARTA information is corresponding to a weekday in March 2020. Fig \ref{fig:R} and Fig \ref{fig:A} show the required energy and the connection time of this day respectively. We also assume that EVs are only be charged one time a day and they need to have the requested energy by the time it disconnect from the grid. The simulation parameters of the EV and the EVA are summarized in Table \ref{EV_EVA_Data}. It almost impossible to choose a single value of $\rho$ that would be suitable for all objectives \cite{mohiti2019decentralized}, so we consider $10^{-3}$ and 1 for EVA's LVM and CCM objectives, respectively. 
The implementations of EV ADMM have been done on a PC with Intel®CoreTM i7 7700 3.60 GHz CPU, 8
cores and 64 GB RAM by MATLAB, using CVX \cite{grant2008cvx} to formulate the problem and GUROBI as the solver \cite{Gurobi}. 
\setcounter{topnumber}{2}
\setcounter{bottomnumber}{2}
\setcounter{totalnumber}{2}
\setcounter{dbltopnumber}{2} 

\begin{table}[hpbt!]
\caption{Simulation Parameters} 
\centering 
\resizebox{1\columnwidth}{!}{%
\begin{tabular}{|c|c|c|} 
\hline 
{symbol} & {Parameter} & {value} \\ [0.5ex]
\hline 
$\overline p^{dis}_{i,t}$, $\overline p^{ch}_{i,t}$ &  Maximal discharging/charging rate  & 8kW \\
$\underline E_{i,t}$ & Minimal energy & 2.5 kWh\\
$\overline E_{i,t}$ & Maximal energy  & 50 kWh\\
$E^{0}_{i}$ & Initial energy  & 2.5 kWh\\
$\eta_{i}^{ch}$ & Charging efficiency & 90\% \\
$\eta_{i}^{dis}$ & Discharging efficiency & 88\% \\
$\alpha$ & Battery depreciation parameter & 0.0125 $\$/kW^{2}$ \\
$\overline p^{dis}_{a,t}$, $\overline p^{ch}_{a,t}$&  Aggregated maximal discharging/charging rate & 136kW \\
[1ex] 
\hline 
\end{tabular}
\label{EV_EVA_Data}
}
\end{table}
\subsection{Simulation Scenarios}
\begin{figure*}[hbpt!]
    \begin{minipage}{1\linewidth}
        \begin{minipage}{0.5\linewidth}
           \centering
           \label{fig:sen1_sub_obj1_first} 
           \includegraphics[width=1\linewidth]{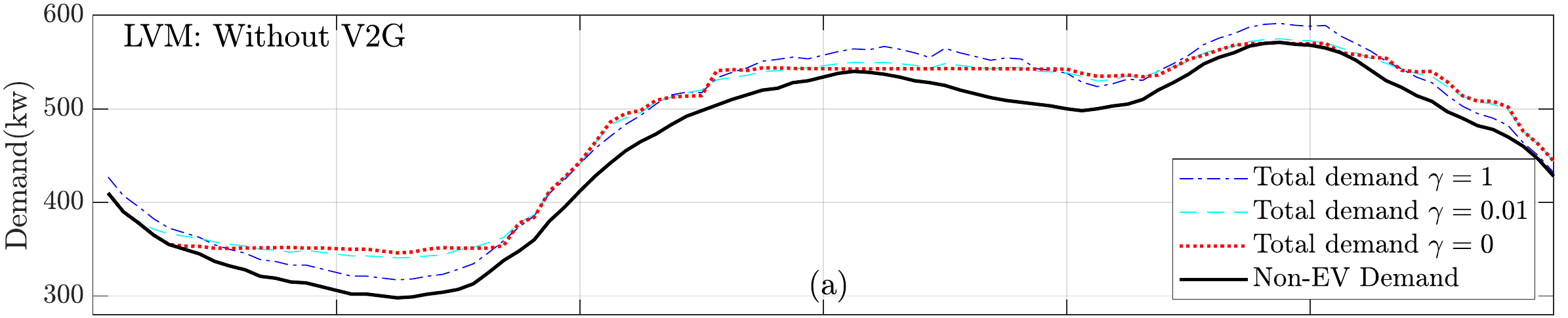}  
        \end{minipage}
        \hfill
        \begin{minipage}{0.5\linewidth}
            \centering
            \label{fig:sen1_sub_obj1_second}
           \includegraphics[width=1\linewidth]{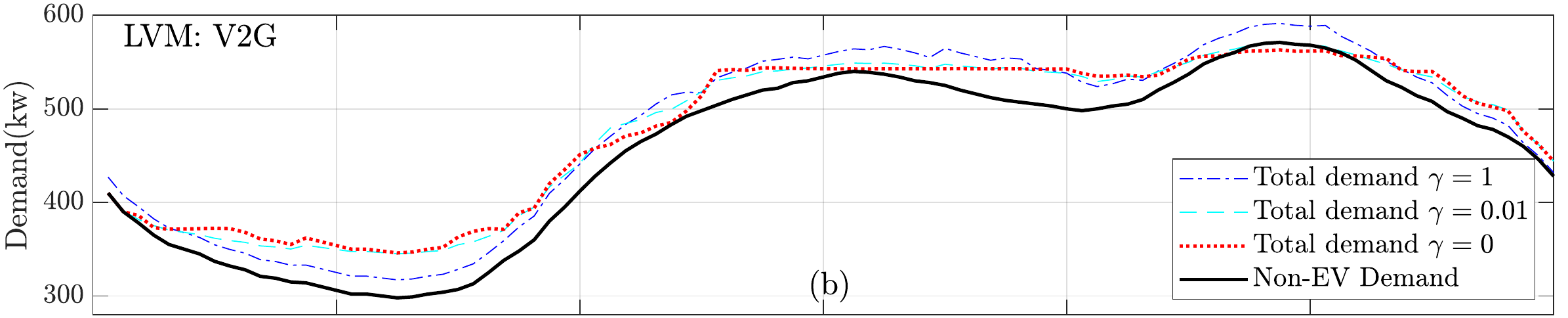}  
        \end{minipage}
        \begin{minipage}{0.5\linewidth}
            \centering
            \label{fig:sen1_sub_obj2_first}
          \includegraphics[width=1\linewidth]{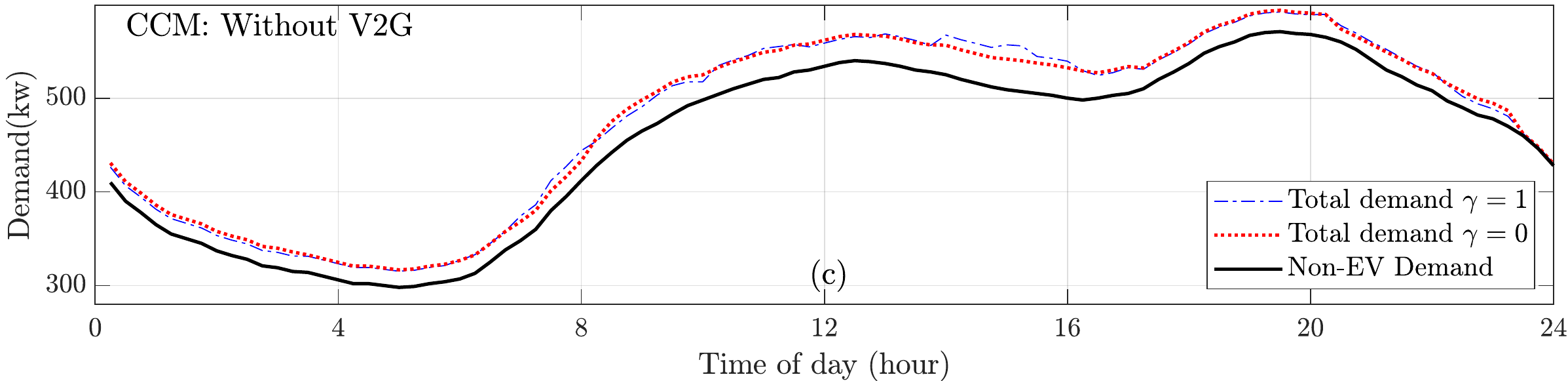} 
        \end{minipage}
          \hfill
        \begin{minipage}{0.5\linewidth}
            \centering
            \label{fig:sen1_sub_obj2_second}
          \includegraphics[width=1\linewidth]{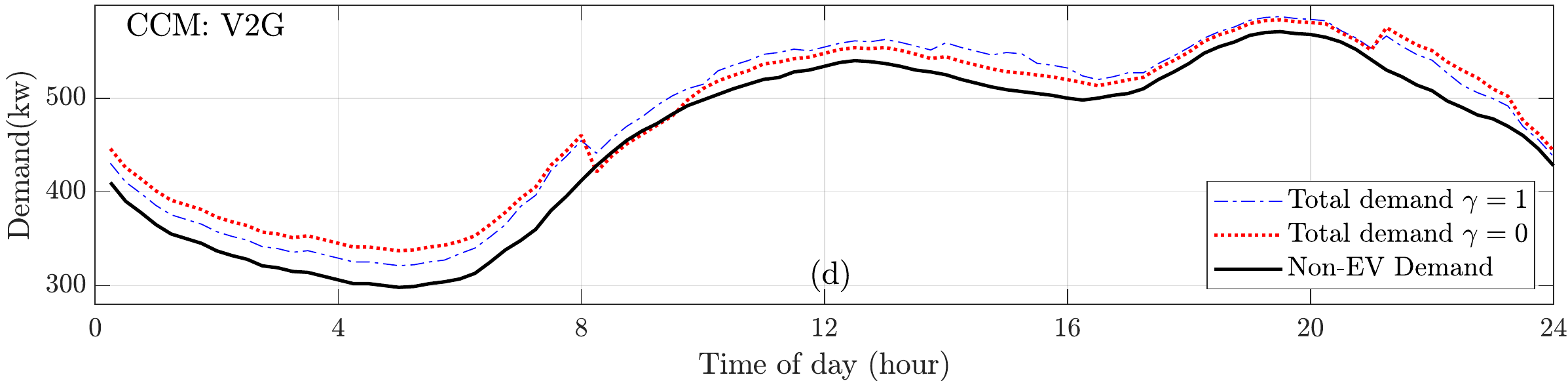} 
        \end{minipage}
        \caption{Scenario 1: (a) LVM: without V2G, (b) LVM: with V2G, (c) CCM: without V2G, (d) CCM: with V2G}
        \label{fig:secn1_compare}
        \begin{minipage}{0.5\linewidth}
            \centering
            \label{fig:sen2_sub_obj1_first}
          \includegraphics[width=1\linewidth]{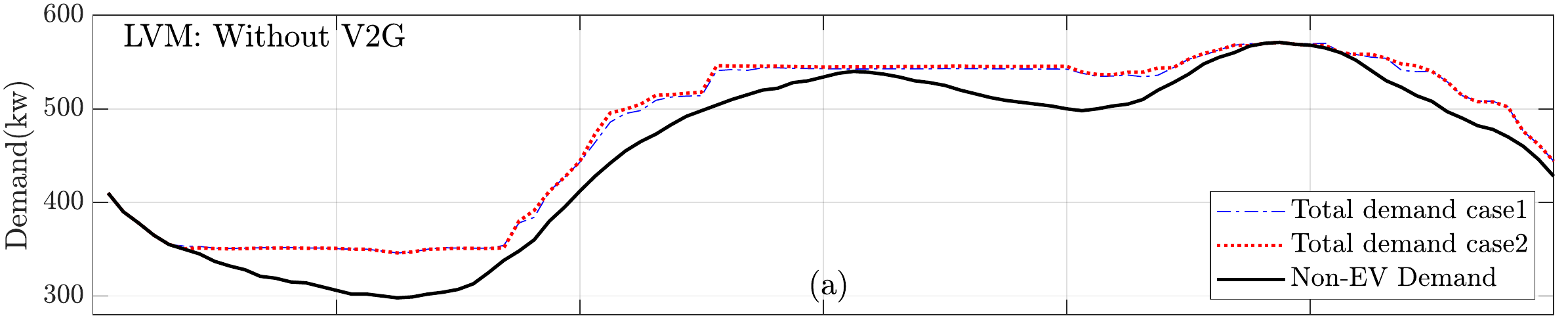} 
        \end{minipage}
          \hfill
        \begin{minipage}{0.5\linewidth}
            \centering
            \label{fig:sen2_sub_obj1_second}
          \includegraphics[width=1\linewidth]{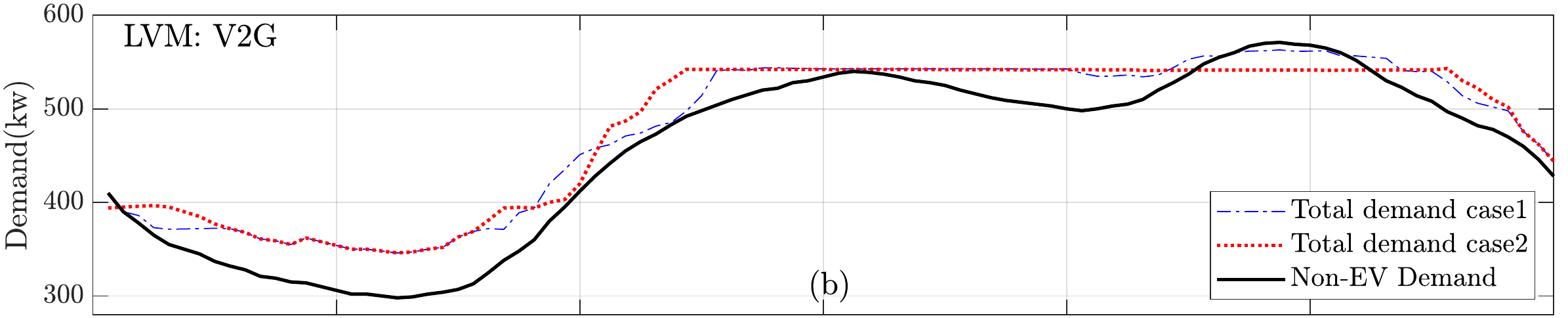} 
        \end{minipage}
        \begin{minipage}{0.5\linewidth}
            \centering
            \label{fig:sen2_obj2_first}
          \includegraphics[width=1\linewidth]{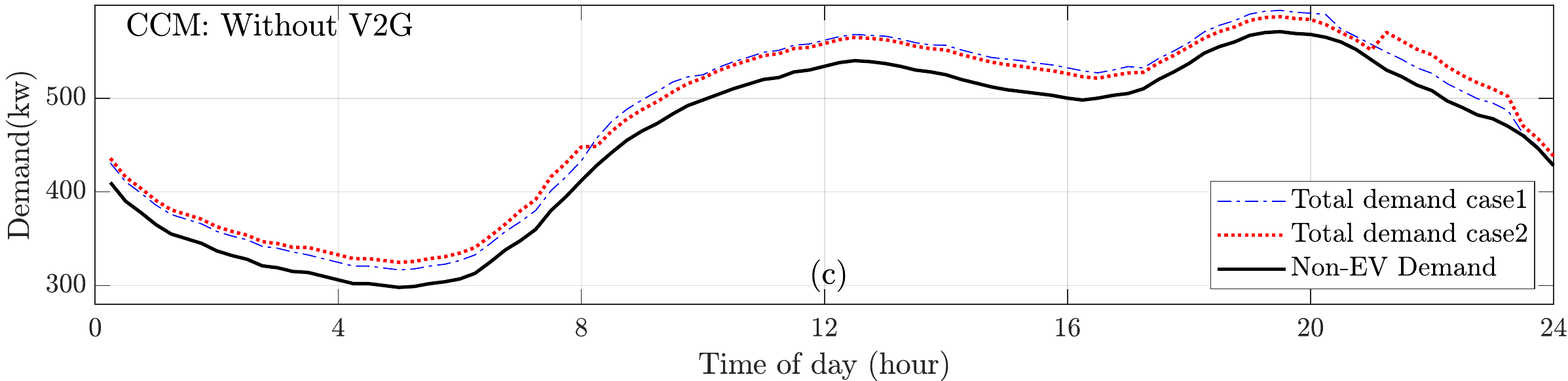} 
        \end{minipage}
          \hfill
        \begin{minipage}{0.5\linewidth}
            \centering
            \label{fig:sen2_obj2_second}
          \includegraphics[width=1\linewidth]{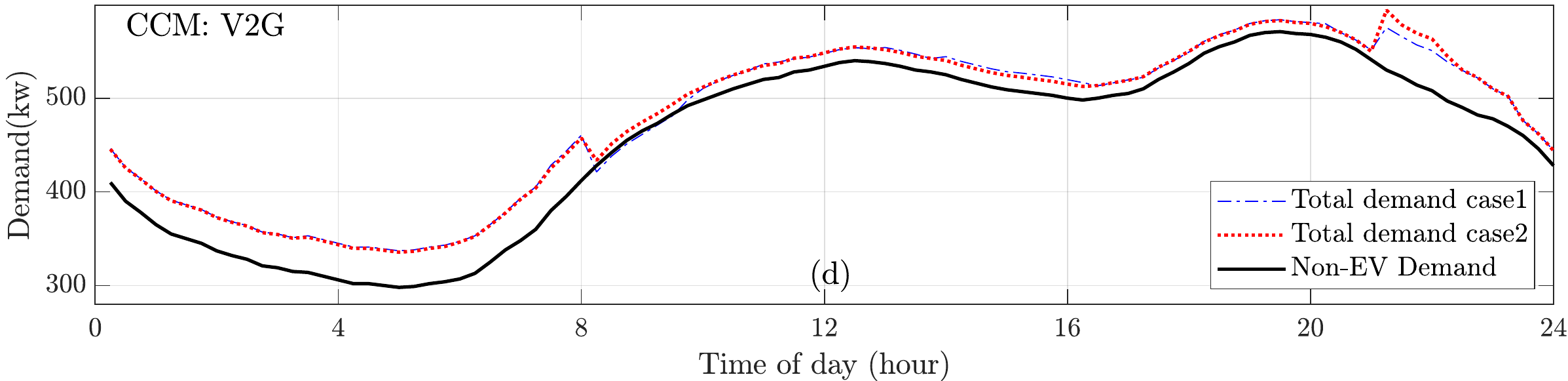} 
        \end{minipage}
        \caption{Scenario 2: (a) LVM: without V2G, (b) LVM: with V2G, (c) CCM: without V2G, (d) CCM: with V2G}
        \label{fig:secn2_compare}
    \end{minipage}
    \\
\end{figure*}

To clearly illustrate the simulation results of the implemented model, two scenarios are considered in this paper. These scenarios are applied to both EVA's objectives as follows:

\textbf{Scenario 1}: In scenario 1, we analyze the impact of $\gamma$ on EVA's objectives. Moreover, we test EVs' discharge capability on both objective functions. Fig \ref{fig:secn1_compare} shows the results for the LVM and CCM. It can be seen from Fig \ref{fig:secn1_compare} (a) and (b) that by increasing $\gamma$, the importance rate of EVs' objective, which is the battery depreciation minimization, increases significantly. Therefore, a high value of $\gamma$ limits extra charging or discharging to fulfill the EVA objective function. Furthermore, compared to the V2G case, the system without V2G is less capable to do valley filling and peak shaving.  

Furthermore, Fig \ref{fig:secn1_compare} (c) and (d) present an improvement in the CCM optimization problem, when EVs have the discharging ability. 

Considering the peak price between 4 pm to 9 pm, Fig \ref{fig:secn1_compare} (d) demonstrates a fluctuation in demand profile at these times. That is because the optimization try to adjust the demand profile to the new electricity price. We can also see that, if battery depreciation cost is considered, the EVs have a low incentive to feed energy back to the grid. On the flip side, the charging profile is almost the same for both $\gamma$ in Fig \ref{fig:secn1_compare} (c). The reason is that based on real EV connection data that we used, EVs are connected only for a certain amount of time steps in a day, so considering the required energy by each EV, the system is not flexible enough to fulfills the EVA objectives.

\textbf{Scenario 2}: To demonstrate the impacts of EV constraints on the final demand profile, we considered two cases in this scenario, where $\gamma$ is equal to zero. In case1, the model is executed considering energy and (dis)charging efficiency constraints for individual EV optimization problems. Different charging and discharging tariffs are also applied to the EVA CCM objective in this case. Conversely, case2 is used as a benchmark, where the model does not have those constraints. It can be seen from Fig \ref{fig:secn2_compare} (a) and (b) that using V2G in case2, the demand profile almost becomes flat in the peak hours, which is not a realistic result even considering the V2G ability. However, in case1, the load variance is not as less the case2. 
    On the other hand, due to the importance of V2G capability in the optimization, in Fig \ref{fig:secn2_compare} (a) both cases follow each other. 
    
    Fig \ref{fig:secn2_compare} (c) and (d) also shows the aggregated charging profile for the cost minimization objective. It is clear from the Fig \ref{fig:secn2_compare} (c) that, applying the new constraints to the model is made the aggregated EVs' demand insensitive to the EVA objective, even when $\gamma=0$. Despite this fact, \ref{fig:secn2_compare} (d) show some changes in the total demand curve, during peak hours, when we the EVs have the V2G ability. It is worth noting that if EVs have more connection time, the difference between these cases will be more, especially when EVs' have the discharging capability.

\begin{figure}[hpbt!]
    \centering
    \includegraphics[width=1\columnwidth]{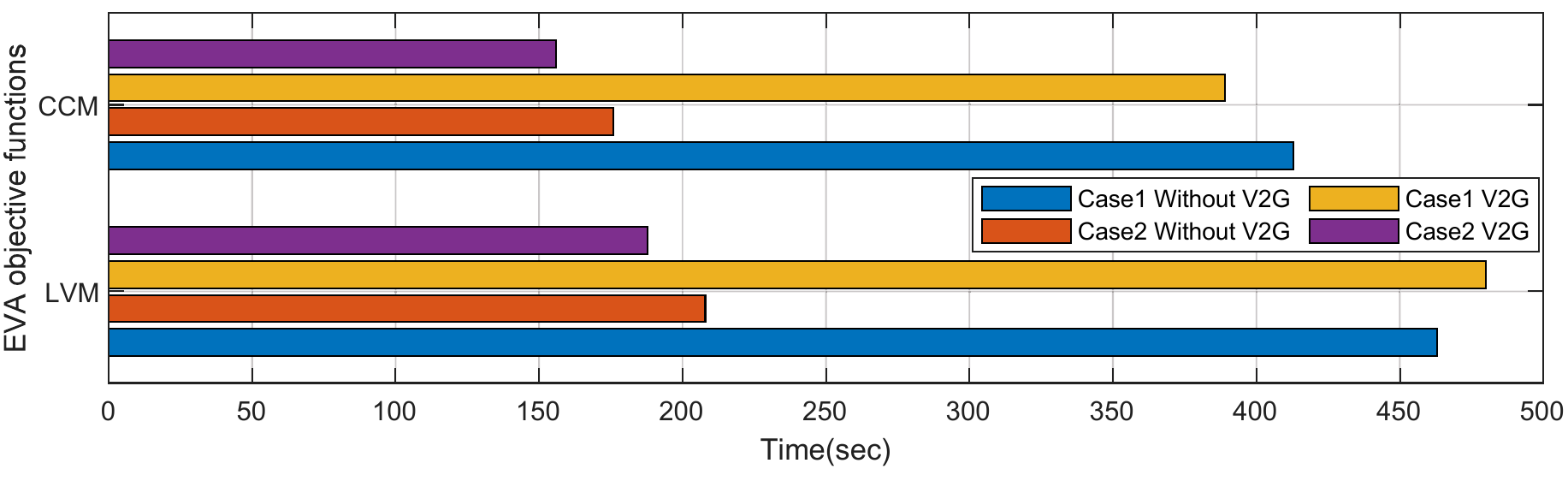}
    \caption{Run-time in scenario 2}
    \label{fig:runtime}
\end{figure}

    Moreover, the simulation results of scenario 2 considering V2G show that compared to the average amount of the total load, load profile becomes less smooth by 12\% for the load variance minimization and 2\% for the cost minimization objectives, when we consider the comprehensive model for battery operation. It worth noting that to implement case1, we need to use an MIQP model, which is much more time consuming compared to case2. Fig \ref{fig:runtime} shows the simulation results for run-time in scenario 2. Considering the MIQP model in case2, run-time, in this case, is more than 2 times of case1 for both EVA objectives. However, there is only a small difference between V2G and without V2G cases in each EVA objective.

\begin{figure}[H]
\centering
    \begin{minipage}{0.8\linewidth}
        \centering
       \includegraphics[width=0.9\linewidth]{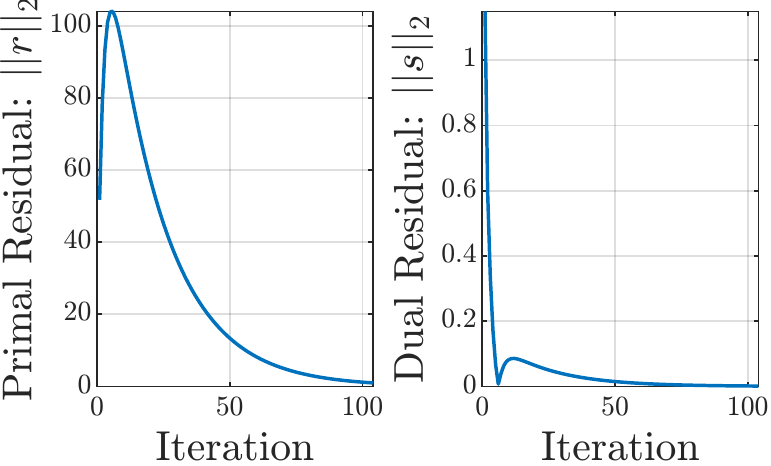}
        \caption{Primal and Dual residuals for LVM optimization}
        \label{fig:residual}
    \end{minipage}
\end{figure}

Note that our proposed ADMM-based method converges to  optimum values as gain $\rho$ for mismatch is less than the critical value. To show the convergence, as an example, Fig \ref{fig:residual} shows  primal and dual residuals for LVM in Scenario 2, case1, converges as the iteration proceeds. The optimum choice of penalty parameter is beyond the scope of this work.

\section{Conclusion}

In this paper, the EVA has two goals, which are load variance minimization and charging cost minimization. Besides, each EV has its optimization problem, which is minimizing the battery depreciation cost. The ADMM-based EV charging problem is formulated as an MIQP optimization problem to demonstrate the impacts of V2G on overall charging cost and the EVA's constraints. Numerical tests with different levels of the importance rate of individual EV goal and EVA objectives indicate that the proposed approach is not only optimal but also scalable. Moreover, the results show the considerable effects of EV energy and efficiency constraints on increasing the standard deviation of the aggregated load profile, and computation time. In future work, we will focus on how can we make the problem fully distributed in the EVA side and consider the satisfaction level on the EV side. Moreover, we plan to use a dynamic penalty parameter to speed up the optimization run-time.

\bibliographystyle{IEEEtran}
\footnotesize
\bibliography{IEEEabrv,Ref}

\begin{thebibliography}{10}
\providecommand{\url}[1]{#1}
\csname url@samestyle\endcsname
\providecommand{\newblock}{\relax}
\providecommand{\bibinfo}[2]{#2}
\providecommand{\BIBentrySTDinterwordspacing}{\spaceskip=0pt\relax}
\providecommand{\BIBentryALTinterwordstretchfactor}{4}
\providecommand{\BIBentryALTinterwordspacing}{\spaceskip=\fontdimen2\font plus
\BIBentryALTinterwordstretchfactor\fontdimen3\font minus
  \fontdimen4\font\relax}
\providecommand{\BIBforeignlanguage}[2]{{%
\expandafter\ifx\csname l@#1\endcsname\relax
\typeout{** WARNING: IEEEtran.bst: No hyphenation pattern has been}%
\typeout{** loaded for the language `#1'. Using the pattern for}%
\typeout{** the default language instead.}%
\else
\language=\csname l@#1\endcsname
\fi
#2}}
\providecommand{\BIBdecl}{\relax}
\BIBdecl

\bibitem{finance2019electric}
B.~N.~E. Finance, \emph{Electric Vehicle Outlook}, 2019,
  \url{https://about.bnef.com/electric-vehicle-outlook/}.

\bibitem{afshar2020literature}
S.~{Afshar}, P.~{Macedo}, F.~{Mohamed}, and V.~{Disfani}, ``A literature review
  on mobile charging station technology for electric vehicles,'' in \emph{2020
  IEEE Transportation Electrification Conference Expo (ITEC)}, 2020, pp.
  1184--1190.

\bibitem{mohsenzadeh2018optimal}
A.~Mohsenzadeh, S.~Pazouki, S.~Ardalan, and M.-R. Haghifam, ``Optimal placing
  and sizing of parking lots including different levels of charging stations in
  electric distribution networks,'' \emph{International Journal of Ambient
  Energy}, vol.~39, no.~7, pp. 743--750, 2018.

\bibitem{wei2015charging}
W.~Wei, F.~Liu, and S.~Mei, ``Charging strategies of ev aggregator under
  renewable generation and congestion: A normalized nash equilibrium
  approach,'' \emph{IEEE Transactions on Smart Grid}, vol.~7, no.~3, pp.
  1630--1641, 2015.

\bibitem{clement2009impact}
K.~Clement-Nyns, E.~Haesen, and J.~Driesen, ``The impact of charging plug-in
  hybrid electric vehicles on a residential distribution grid,'' \emph{IEEE
  Transactions on power systems}, vol.~25, no.~1, pp. 371--380, 2009.

\bibitem{zheng2018novel}
Y.~Zheng, Y.~Shang, Z.~Shao, and L.~Jian, ``A novel real-time scheduling
  strategy with near-linear complexity for integrating large-scale electric
  vehicles into smart grid,'' \emph{Applied Energy}, vol. 217, pp. 1--13, 2018.

\bibitem{kia2019tutorial}
S.~S. Kia, B.~Van~Scoy, J.~Cortes, R.~A. Freeman, K.~M. Lynch, and S.~Martinez,
  ``Tutorial on dynamic average consensus: The problem, its applications, and
  the algorithms,'' \emph{IEEE Control Systems Magazine}, vol.~39, no.~3, pp.
  40--72, 2019.

\bibitem{yang2019survey}
T.~Yang, X.~Yi, J.~Wu, Y.~Yuan, D.~Wu, Z.~Meng, Y.~Hong, H.~Wang, Z.~Lin, and
  K.~H. Johansson, ``A survey of distributed optimization,'' \emph{Annual
  Reviews in Control}, vol.~47, pp. 278--305, 2019.

\bibitem{disfani2015optimization}
V.~R. Disfani, \emph{Optimization and control for microgrid and power
  electronic converters}.\hskip 1em plus 0.5em minus 0.4em\relax University of
  South Florida, 2015.

\bibitem{mohammadi2016fully}
J.~Mohammadi, G.~Hug, and S.~Kar, ``A fully distributed cooperative charging
  approach for plug-in electric vehicles,'' \emph{IEEE Transactions on Smart
  Grid}, vol.~9, no.~4, pp. 3507--3518, 2016.

\bibitem{gan2013optimal}
L.~Gan, U.~Topcu, and S.~H. Low, ``Optimal decentralized protocol for electric
  vehicle charging,'' \emph{IEEE Transactions on Power Systems}, vol.~28,
  no.~2, pp. 940--951, 2013.

\bibitem{liu2019decentralized}
M.~Liu, P.~K. Phanivong, Y.~Shi, and D.~S. Callaway, ``Decentralized charging
  control of electric vehicles in residential distribution networks,''
  \emph{IEEE Transactions on Control Systems Technology}, vol.~27, no.~1, pp.
  266--281, 2019.

\bibitem{boyd2011distributed}
S.~Boyd, N.~Parikh, E.~Chu, B.~Peleato, and J.~Eckstein, ``Distributed
  optimization and statistical learning via the alternating direction method of
  multipliers,'' \emph{Foundations and Trends{\textregistered} in Machine
  learning}, vol.~3, no.~1, pp. 1--122, 2011.

\bibitem{mota2011proof}
J.~F. Mota, J.~M. Xavier, P.~M. Aguiar, and M.~P{\"u}schel, ``A proof of
  convergence for the alternating direction method of multipliers applied to
  polyhedral-constrained functions,'' \emph{arXiv preprint arXiv:1112.2295},
  2011.

\bibitem{wasti2020distributed}
S.~Wasti, P.~Ubiratan, S.~Afshar, and V.~Disfani, ``Distributed dynamic
  economic dispatch using alternating direction method of multipliers,''
  \emph{arXiv preprint arXiv:2005.09819}, 2020.

\bibitem{vaya2014decentralized}
M.~G. Vay{\'a}, G.~Andersson, and S.~Boyd, ``Decentralized control of plug-in
  electric vehicles under driving uncertainty,'' in \emph{IEEE PES Innovative
  Smart Grid Technologies, Europe}.\hskip 1em plus 0.5em minus 0.4em\relax
  IEEE, 2014, pp. 1--6.

\bibitem{zhang2017scalable}
L.~Zhang, V.~Kekatos, and G.~B. Giannakis, ``Scalable electric vehicle charging
  protocols,'' \emph{IEEE Transactions on Power Systems}, vol.~32, no.~2, pp.
  1451--1462, 2017.

\bibitem{carli2017decentralized}
R.~Carli and M.~Dotoli, ``A decentralized control strategy for optimal charging
  of electric vehicle fleets with congestion management,'' in \emph{2017 IEEE
  International Conference on Service Operations and Logistics, and Informatics
  (SOLI)}.\hskip 1em plus 0.5em minus 0.4em\relax IEEE, 2017, pp. 63--67.

\bibitem{rivera2017distributed}
J.~Rivera, C.~Goebel, and H.-A. Jacobsen, ``Distributed convex optimization for
  electric vehicle aggregators,'' \emph{IEEE Transactions on Smart Grid},
  vol.~8, no.~4, pp. 1852--1863, 2017.

\bibitem{mohiti2019decentralized}
M.~Mohiti, H.~Monsef, and H.~Lesani, ``A decentralized robust model for
  coordinated operation of smart distribution network and electric vehicle
  aggregators,'' \emph{International Journal of Electrical Power \& Energy
  Systems}, vol. 104, pp. 853--867, 2019.

\bibitem{khaki2019hierarchical}
B.~Khaki, C.~Chu, and R.~Gadh, ``Hierarchical distributed framework for ev
  charging scheduling using exchange problem,'' \emph{Applied energy}, vol.
  241, pp. 461--471, 2019.

\bibitem{miao2013soc}
Z.~Miao, L.~Xu, V.~R. Disfani, and L.~Fan, ``An soc-based battery management
  system for microgrids,'' \emph{Ieee transactions on smart grid}, vol.~5,
  no.~2, pp. 966--973, 2013.

\bibitem{feizollahi2015large}
M.~J. Feizollahi, M.~Costley, S.~Ahmed, and S.~Grijalva, ``Large-scale
  decentralized unit commitment,'' \emph{International Journal of Electrical
  Power \& Energy Systems}, vol.~73, pp. 97--106, 2015.

\bibitem{takapoui2017alternating}
R.~Takapoui, ``The alternating direction method of multipliers for
  mixed-integer optimization applications,'' Ph.D. dissertation, Ph. D.
  dissertation, Stanford University, 2017.

\bibitem{munich}
\BIBentryALTinterwordspacing
\emph{SWM München. Netzdaten.}, (accessed Apr 26, 2020). [Online]. Available:
  \url{https://www.swm-infrastruktur.de/strom/netzstrukturdaten/netzdaten.html}
\BIBentrySTDinterwordspacing

\bibitem{carta}
\BIBentryALTinterwordspacing
\emph{Chattanooga Area Regional Transportation Authority}, (accessed Apr 26,
  2020). [Online]. Available: \url{http://www.carta-bus.org/}
\BIBentrySTDinterwordspacing

\bibitem{tou}
\BIBentryALTinterwordspacing
\emph{Time-Of-Use (TOU) Rate Plans}, (accessed Apr 26, 2020). [Online].
  Available:
  \url{https://www.sce.com/residential/rates/Time-Of-Use-Residential-Rate-Plans}
\BIBentrySTDinterwordspacing

\bibitem{grant2008cvx}
M.~Grant, S.~Boyd, and Y.~Ye, ``Cvx: Matlab software for disciplined convex
  programming,'' 2008.

\bibitem{Gurobi}
\BIBentryALTinterwordspacing
\emph{Gurobi Optimization}. [Online]. Available: \url{http://www.gurobi.com}
\BIBentrySTDinterwordspacing

\end{thebibliography}

\end{document}